%% file: paper.tex
\input{feynman}
\documentstyle[aps,preprint,psfig]{revtex}
\tighten
\begin{document}  
\preprint{TAUP-2328-96}
\title{\bf Feynman Graphs and Generalized Eikonal Approach to 
High Energy Knock-Out Processes}
\author{L.~L.~Frankfurt$^{a,c}$, M.~M.~Sargsian$^{a,d}$ , 
M.~I.~Strikman$^{b,c}$ } 
\address{
$(a)$ School of Physics and Astronomy, Tel Aviv University, Tel Aviv, 
69978, Israel \\
$(b)$ Department of Physics, 
Pennsylvania State University, University Park, PA 16802\\
$(c)$ Institute for Nuclear Physics, St. Petersburg, Russia \\
$(d)$ Yerevan Physics Institute, Yerevan, 375036, Armenia } 
\maketitle
{\centerline{\today{}}
\begin{abstract}
The cross section of hard semi-exclusive $A(e,e'N)(A-1)$ reactions for
fixed missing energy and momentum is calculated within the eikonal 
approximation. Relativistic dynamics and kinematics of high energy 
processes are unambiguously accounted for by using the analysis of 
appropriate Feynman diagrams. A significant dependence of the 
final state interactions on the missing energy is found, which is 
important  for interpretation of forthcoming color transparency 
experiments.  
A new, more  stringent kinematic restriction on the region 
where the contribution of  short-range nucleon correlations is 
enhanced in semi-exclusive knock-out processes is derived.
It is also demonstrated that the use of light-cone variables 
leads to a considerable simplification of the  description of 
high-energy knock-out reactions.

\end{abstract}


\section{introduction}

With the advance of high energy,  high intensity electron facilities 
(see e.g. \cite{CEBAF,HERMES,ELFE})  the high momentum transfer 
semi-exclusive reactions are  becoming a practical tool for the 
investigation of the microscopic structure of nuclei, nuclear matter
and the color transparency phenomenon.
However, theoretical methods which were successful in  medium-energy  
nuclear physics should be upgraded in order to describe processes  where  
energies transferred to a nuclear target are $\gtrsim few$ $GeV$. 
This paper  focuses on the calculation of the influence of the 
final state interactions (FSI) on high energy hard semi-exclusive
$A(e,e'N)X$ reactions, for  energies of knocked out nucleon
$E_N \ge 1 GeV$ and for states $X$ 
representing  ground or excited states of the residual nucleus.

At energies $E_N \le 1 GeV$ the final state interactions (FSI)
are usually evaluated in terms of interactions of knock-out nucleons
with  an effective potential of the residual system -
the optical model approximation, (see, for example, Ref.\cite{OV}). 
Parameters of the effective potential are 
adjusted to describe data on elastic
$N-(A-1)$ scattering  for  projectile energies close to $E_N$.
Two important features of high-energy FSI  make the extension of 
the medium energy formalism to high energies problematic. 
Firstly, the number of essential partial waves increases rapidly 
with the energy of the $N,(A-1)$ system. Secondly, the $NN$ interaction 
which is practically elastic for $E_N \le 500 MeV$ becomes predominantly 
inelastic for $E_N > 1 GeV$. 
Hence the problem of scattering hardly can be treated 
as a many body quantum mechanical problem. Introducing
in this situation a predominantly imaginary potential to 
account for hadron production (not
only for excitations of residual system as in the case of intermediate
energies) is not a  well defined mathematical concept. 
So theoretical methods successful below 1 GeV become ineffective at the 
energies which can be probed  at  Jefferson Lab \cite{CEBAF}, 
HERMES\cite{HERMES} and  ELFE\cite{ELFE}.

FSI at higher energies ($E_N>1~GeV$) are often described within the 
approximation of the additivity of phases,  acquired in the sequential
rescatterings of high-energy projectiles off the target nucleons 
(nonrelativistic Glauber model \cite{Glauber}). This approximation
made it possible to describe the data on elastic $hA$ scattering 
at hadron energies  $1~GeV <E_h <10-15~GeV$ (cf. Refs. \cite{Yennie,Moniz}). 
It has been also  applied  to  the description of cross sections of 
$A(e,e'N)(A-1)$ reactions\cite{F1,M1,M2,Si,Zv,Br,R2,N1,Wt,Bi,MP} integrated over 
the excitation  energies ($E_{exc}$) of the residual $(A-1)$ nucleon 
system, for small  momentum of the residual system $\vec p_{A-1} \le p_F$.
In Ref.\cite{FSZ,FMSS95} the cross section of $A(e,e'N)(A-1)$ reactions 
has been calculated for  small excitation  energies that are 
characteristic for particular shells of a target nucleus with 
$A\gtrsim 12-16$. Thus the dependence of FSI on missing energy-$E_m$ 
($\sim E_{exc}$) was not essential in the previous calculations. 
Furthermore such a dependence  is not important for total cross 
sections, small angle coherent and noncoherent (summed over residual  
nuclear excitations) scatterings in $hA$ reactions. 
However the dependence of FSI on the missing energy is part 
of the color transparency phenomena in high-energy quasielastic  
processes where restrictions on the missing energy should be 
imposed to suppress inelastic processes where pions are produced
\cite{Mueller,Brodsky}. 
It is also important  in the studying of  short-range nucleon 
correlations in nuclei in  semi-exclusive reactions, where large 
value of missing energy should be ensured.
 
In this paper we consider high-energy semi-exclusive $A(e,e'N)(A-1)$ 
reactions, where both missing momentum and missing energy are 
fixed. We investigate the implications of the nonzero value 
of the  missing energy on the FSI  of the knocked-out nucleon.

The linear increase with incident energy of the coherence length 
of strong interactions leads to a change of the underlying
physical picture of hadron-nucleus scattering - from sequential 
rescatterings to coherent interactions with all nucleons at a 
given impact parameter. First evidence for that was obtained 
by Mandelstam\cite{Man} who analyzed planar  Feynman diagrams 
corresponding to rescattering diagrams of the nonrelativistic  
Glauber approximation. He found that the contribution of these  
diagrams tends to zero in the high-energy limit.

Later on Gribov\cite{GIN} developed a quantitative theory of
high-energy $hA$ interactions and demonstrated that the small 
value of  the ratio of inelastic diffractive and elastic  
cross sections serves as a small parameter, justifying  
Glauber-type formulae. We restrict the analysis in the paper 
to the range  of energies where inelastic diffraction 
in the soft hadron processes is a small correction 
(i.e. energies  of the  knocked-out nucleon $\lesssim 10~GeV$).
We also restricted by the photon virtualities  $Q^2 \sim 1-3~GeV^2$,  
where color coherent phenomena are expected to be a small correction.
At larger $Q^2$ the deduced formulae can be used as a  baseline model
for searching for  color coherent phenomena. (Note that the  small 
ratio of inelastic and elastic diffraction reflects small dispersion  
of strengths of interaction for soft processes. These fluctuations 
are naturally much larger in the case of hard processes where a probe 
selects a rare, small size configurations in the struck  
nucleon. For a review of physics of the color coherence phenomena 
see Refs.\cite{FMS94,JPR96}). 

At large $Q^2$ and small struck nucleon momenta it is safe to neglect 
the dependence of the $e N$ scattering amplitude on nucleon binding 
since the energy scale of the hard interaction is much larger than  
the nuclear energy scale. At the same time, when missing momenta 
and missing energies relevant for  knock-out processes are comparable
with the Fermi momenta and the nucleon binding energy, it is necessary
to take them into account in the calculation of the nuclear part of  
the scattering amplitude. Obviously, this can not be done unambiguously 
within the optical model and the Glauber type approximations, which  
neglect nucleon Fermi momenta in the nuclei. To calculate FSI of the 
knocked out nucleon we derive the formulae of the eikonal approximation  
which account for the nucleon Fermi motion. Our derivation  is based
on  the analysis of the Feynman diagrams corresponding to the $A(e,e'N)X$ 
reaction.

The method of the derivation of the formulae of the eikonal approximation  
on the basis of the analysis of the relevant  Feynman diagrams has been 
suggested long ago for hadron-nucleus collisions in 
Refs.\cite{Gribov,Bertocchi}. It has been shown in Refs.\cite{Gribov,Bertocchi}    
under what conditions the Feynman diagram description of the hadron-nucleus 
scattering processes leads to the optical model or the Glauber type 
approximation. The main advantage of the Feynman graph approach is that  
it takes into account the relativistic kinematics  of high-energy processes.
Particularly it accounts for an important feature of high energy small-angle
elastic (diffractive) scatterings -- the conservation of the light-cone  
momentum  $p_-\equiv\sqrt{m^2+p^2}-p_z$, where $p_z$ is the  component of 
nucleon Fermi momentum $\vec p$ in the projectile momentum direction. 
In the present paper we apply this  method  to calculate FSI in 
$A(e,e'N)(A-1)$ reactions  and  extend the results of ref.\cite{FGMSS95}  
for $^2H(e,e')(pn)$ process to the case of nucleon knock-out processes
off $^3He$($^3H$) (eq.(\ref{amp_sum3})). After  deriving formulae of the 
impulse approximation, single and double rescattering terms,  
we generalize the obtained results to  the case of a nucleus 
with arbitrary $A$. In the case of a deuteron target we found  
significant effects of the missing energy\cite{FGMSS95} (Fig.3). 
We demonstrate that even larger effects are expected for knock 
out of nucleons off $^3He$,$^4He$ targets.

It follows from the formulae derived in the paper that when missing 
momenta and energy are  not negligible in knock-out reactions,
the optical model approximation becomes unreliable. 
Also, we found in this kinematics significant corrections to the 
conventional  Glauber type formulae (Fig.4).

Based on the analysis of the derived formulae we determine
optimal kinematic conditions  for the investigation of the 
short-range nucleon correlations in nuclei in semi-inclusive  
reactions. In particular, if one wants to use for such 
investigations kinematics $x_{Bj}>1$  it is necessary
to impose additional conditions on the recoil energy of the 
residual system (eq.(40)). Such  conditions allow  
to suppress the contribution of low-momentum component
of the nuclear wave function due to  FSI (Section 4).

We demonstrate also that light-cone kinematics of high-energy 
knock-out reaction is naturally accounted for if nucleon Fermi 
momenta in the nucleus are parameterized in terms of the 
light-cone variables.

\section{Scattering amplitude}

In this section we  consider the scattering amplitude for a knocked-out  
nucleon  to  undergo  $n$ rescatterings off the  nucleons of $(A-1)$  
residual system. The case $n=0$ corresponds to the impulse approximation  
(IA) in which the knocked out nucleon does not interact with residual
nucleus. We systematically neglect in this paper the diffractive  
excitation of the nucleons in the intermediate states. In soft QCD  
processes this is a small correction for the knock-out nucleon energies  
$\lesssim 10~GeV$. In the hard processes (that is when $Q^2$ - virtuality  
of the photon is sufficiently large ($\gtrsim 6-8~GeV^2$)) such an 
approximation can not be justified even  within this energy range, 
see for example discussion in  Ref.\cite{FMS94}. However our aim is  
to perform calculations in the kinematics where CT phenomenon is still 
a small correction.

The  scattering amplitude can be represented by covariant 
Feynman diagrams of Fig.1, in the approximation when only 
elastic rescatterings are accounted for, as:

\centerline{
\input{figure2.tex}
}

\vspace{0.8cm}

\noindent {\bf Fig. 1} 
\ {\em $n$-fold rescattering diagram.}
 
\vspace{0.5cm}

\begin{eqnarray}
& & F^{(n)}_{A,A-1}(q,p_f) = 
\sum\limits_{h}{1 \over n!(A-n-1)!
}\prod\limits_{i=1}^{A}
\prod\limits_{j=2}^{A} 
\int d^4p_i d^4p'_j 
{1\over \left[ i (2\pi)^4\right]^{n+1}}
\nonumber \\
& & 
\delta^4(\sum\limits_{i=1}^{A}p_i-{\cal P}_A) 
\delta^4(\sum\limits_{j=2}^{A}p'_j-{\cal P}_{A-1}) 
\prod\limits_{m=n+2}^{A}\delta^4(p_m-p'_m) \times 
\nonumber \\
& &  
{\Gamma_{A}(p_1,...,p_A)\over D(p_1)D(p_2)..D(p_{n+1})D(p_{n+2})..D(p_A)} 
{F^{em}_h(Q^2)\over  D(p_1+q)} {f^{NN}_1(p_2,p'_2)
  .. f^{NN}_n(p_{n+1},p'_{n+1}) 
\over D(l_1)..D(l_k).. D(l_{n-1})}
\times \nonumber \\       
& & {\Gamma_{A-1}(p'_2,.p'_{n+1},.p_{n+2}..,p_A)\over 
D(p'_2)..D(p'_{n+1})} 
\label{amp_n} 
\end{eqnarray}
where, for the sake of simplicity, we  neglected the spin dependent  
effects. Here  ${\cal P}_A$ and  ${\cal P}_{A-1}$ are the four  momenta 
of target nucleus, and  final $(A-1)$ system, $p_j$ and $p'_j$  
are nucleon momenta in the nucleus $A$ and residual $(A-1)$ system 
respectively. $\sum\limits_{h}$ in eq.(\ref{amp_n}) goes over virtual
photon  interactions  with different nucleons, where  $F^{em}_h(Q^2)$
is electromagnetic vertices. 
$-D(p_k)^{-1}=(p_k^2-m^2+i\epsilon)^{-1}$ 
is  the propagator  of a nucleon with momentum $p_k$. 
The  $f^{NN}_k(p_{k+1},p'_{k+1})$ is the amplitude  of $NN$ scattering,  
${d \sigma \over d t}\sim{|f|^2\over s_k^2}$, where $s_k$ is the total invariant 
energy of two interacting nucleons. $-D(l_k)^{-1}$ is the propagator 
of the struck nucleon in the intermediate  state, with momentum 
$l_k=q+p_1+\sum\limits_{i=2}^{k}(p_i-p'_i)$  between ${k-1}$-th and $k$-th  
rescatterings.  The factor $n!(A-n-1)!$ accounts for  the combinatorics  
of $n$- rescatterings and $(A-n-1)$ spectator nucleons. Following  
Ref.\cite{Gribov}  we choose the ``minus'' sign for the nucleon 
propagators to simplify the calculation of the overall sign of 
the scattering  amplitude - for each closed contour one gets 
the factor  ${1\over i(2\pi)^4}$   with no additional sign.

\centerline{ 
\input{figure3.tex}}

\vspace{8.5cm}

\noindent {\bf Fig. 2} 
\ {\em   Feynman diagrams corresponding to  $^3He(e,e'p)pn$   scattering.  
Dashed lines represent effective  $NN$ scattering,  full  circle represents 
the residual  interaction  between  spectator nucleons.}

 \vspace{0.8cm}

The vertex functions $\Gamma_{A}(p_1,...,p_A)$ and 
$\Gamma_{A-1}(p'_2,...,p_A)$ describe  transitions of 
$''nucleus~ A''$ to $''A~nucleons''$ with momenta $\{p_n\}$ 
and transitions of $''(A-1)~nucleons''$ with momenta $\{p'_n\}$ 
to $``(A-1)~ nucleon~final ~state''$ respectively.
The intermediate spectator state in the diagram of Fig.1 
is expressed in terms of nucleons because the closure over  
various nuclear excitations in the intermediate state is 
used (for the details see Appendix A).

After evaluation of the intermediate state nucleon propagators the  
covariant amplitude  will be reduced to  a set of time ordered   
noncovariant diagrams. This will help to  establish the correspondence
between the vertex functions and the nuclear wave functions. We derive
formulae for the impulse approximation and first two rescattering   
terms (i.e. single and double rescattering). To simplify  derivations 
we consider  $(e,e'N)$ reactions off a three nucleon system  (see Fig.2)  
and then  generalize obtained results to an  arbitrary $A$.

\section{Impulse approximation}
\label{IA}

First, we consider the  $A(e,e'N)(A-1)$ reaction, where the final state 
consists of a noninteracting energetic nucleon $N$ and a $(A-1)$ residual  
state  which can be either the nuclear bound state or break up system 
of  $(A-1)$ nucleons. For the scattering off a three nucleon system this 
reaction corresponds to the covariant diagram of Fig.2a, which is the 
$n=0$ term in the scattering amplitude of eq.(\ref{amp_n}). For the $n=0$ 
term of eq.(\ref{amp_n}), performing the integrations over 
$\delta$-functions due to the energy-momentum conservation we obtain:
\begin{equation}
F^{(a)}\equiv F^{(0)}_{A,A-1}(q,p_f) = 
\int d^4p_3 {\Gamma_A(p_1,p_2,p_3)F^{em}_1(Q^2)\Gamma(p_2,p_3)\over 
D(p_1)D(p_2)D(p_3)},
\label{amp_0_1}
\end{equation}
where
\begin{eqnarray}
& & p_1  =  {\cal P}_{A}-{\cal P}_{A-1},\nonumber \\
& & p_2 + p_3  =  {\cal P}_{A-1}.
\label{kin0}
\end{eqnarray}
Here $\Gamma_A$ and $\Gamma_{A-1}$ correspond to the  nuclear vertices 
represented in Fig.2, by empty and full circles respectively.
In eqs.(\ref{amp_0_1}) and (\ref{kin0}) $p_1$ is the momentum of the 
interacting nucleon and $p_2$, $p_3$ are the momenta of spectator nucleons.
To  simplify the derivation we neglect  the antisymmetrization of
the initial and final nucleon states, which can be easily 
accounted for through the corresponding wave functions (see below).

The scattering amplitude $F^{(a)}$ is Lorenz invariant and corresponds to 
the sum of the non-covariant diagrams with different time orderings 
between  nuclear $\Gamma_A$, $\Gamma_{A-1}$ and electromagnetic vertices 
$F^{em}_h$.  The impulse approximation corresponds to the time ordered 
noncovariant diagram of Fig.2a, where virtual photon is absorbed  by a  
target nucleon which does not interact in the final state. Other time
orderings  correspond to vacuum fluctuations.}

We will perform calculation in the nucleus rest frame in the 
kinematics where  Fermi momenta of target nucleons are not large.
Hence we will restrict the consideration to the range of missing momenta 
$p_m$ and missing energies $E_m$: 
\begin{eqnarray}
|\vec p_m| &  \equiv & |\vec p_f - \vec q| \lesssim 400~MeV/c  \nonumber \\
\alpha & \equiv & {E_f-p_{fz}\over m}- {q_o-|\vec q|\over m}\approx  
{m-E_m - p_{mz}\over m} \approx 1\pm 0.3,
\label{kin1}
\end{eqnarray}
where $\vec p_f$ is the momentum of final knocked-out nucleon, $q_0$  
and $\vec q$ are the energy and momentum of  virtual photon. 
The $z$ axis is chosen in the $\vec q$ direction. The direction 
transverse to  the $\vec q$ would be  labeled by $t$.
$E_m\equiv E_{A-1}+m - E_A$. $\alpha$ is the light-cone fraction of 
the momentum of the target carried by interacting  nucleon
scaled to vary between 0 and $A$.

If we restrict by the kinematics defined in eq.(\ref{kin1}) 
then in the set of the noncovariant diagrams, comprising 
the covariant diagram of Fig.2a, one can neglect the    
diagrams which correspond to the vacuum fluctuations,
(see e.g. Refs.\cite{FS81}).
The latter  become increasingly important  at  larger Fermi momenta  
of the target nucleons. An effective method to account for the diagrams 
with vacuum fluctuations is light-cone approach \cite{FS81,FS88,FS91,K1,K2}  
where  for some components  of the electromagnetic current 
("good" components) their contribution  is suppressed and scattering 
amplitude has the form rather similar to the conventional impulse 
approximation. This physics, being interesting by itself, is beyond 
of the scope of this paper.

Overall in the discussed kinematics (eq.(\ref{kin1})) the relativistic
effects in the nuclear  wave function  are a small correction 
\cite{FS81,FGMSS95} and the  impulse approximation can be calculated 
via nonrelativistic reduction of the covariant nuclear vertices in  
eq.(\ref{amp_0_1}). Such a reduction corresponds to  taking the residue 
over $d p^0_3$, at the nearest nucleon pole in the spectator nucleon  
propagator $D(p_3)^{-1}$. Thus we neglect nonnucleon degrees of freedom 
in a nucleus. The restriction by the nearest  pole in the nucleon
propagators follows from the observation that in the considered 
kinematics (\ref{kin1}) where nuclear excitations are small as compared 
to the scale of energies  characteristic for the  nucleon excitations, 
this is the only pole not corresponding to $N\bar N$ production.
Neglect of discontinuities related to the thresholds of pion production
is justified in QCD  because, for a small pion momenta, the  
pions are Goldstone bosons of spontaneously broken chiral 
symmetry (see the discussion in Ref.\cite{FS88}). 

Taking residue over the spectator nucleon propagator effectively 
corresponds to the replacement 
$$\int  {dp^0_3\over 2\pi i}{1\over D(p_3)} 
\rightarrow {1\over E_3}\approx {1\over 2m}.$$
After this integration one is left with the time ordered diagram 
corresponding to the IA, where virtual photon knocks-out the target nucleon 
with momentum $p_1$,  leaving the residual $A-1$ nucleus with a particular 
excitation energy $E_{exc} = E_m - {p^2_m\over M_{A-1}} - |\epsilon_A|$ where 
$\epsilon_A$ - is the binding energy of the target nucleus.
Nonrelativistic reduction allows to define the momentum space wave function 
through the vertex function  as (c.f. Refs.\cite{Gribov,Bertocchi}):
\begin{equation}
\psi_{A}(p_1,p_2,...p_A)   =   {1\over (\sqrt{(2\pi)^3 2m})^{A-1}}
{\Gamma_A(p_1,p_2,...p_A)\over D(p_1)}, 
\label{wf_p} 
\end{equation}
where   wave functions are normalized as:   
$\int|\psi_A(p_1,p_2,...p_A)|^2 d^3p_1d^2p_2.. d^3p_A = 1$. 
We define the wave function of the final 
(residual nucleus +  knocked-out nucleon) state 
as $\psi_{A-1}/\sqrt{2m}$, where $\psi_{A-1}$ defined 
according to eq.(\ref{wf_p}), with $A$ replaced by $A-1$ 
and the additional factor $1/\sqrt{2m}$ accounts for 
the normalization of the knocked-out nucleon wave function. 
With these definitions  eq.(\ref{amp_0_1}) 
obtains the form of the conventional IA expression:
\begin{equation}
T^{(a)} = \sqrt{(2\pi)^3}(2\pi)^3
\int d^3p_3
\psi_A(p_m,p_2,p_3)F^{em}_1(Q^2)\psi^+_{A-1}(p_2,p_3),
\label{amp_0_2}
\end{equation}
where $\vec p_m = \vec p_f - \vec q$ - is the measured missing momentum,
and $\vec p_2 = -\vec p_3 - \vec p_m$. 
The spin and isospin indices and antisymmetrization of wave functions 
are implicit in eq.(\ref{amp_0_2}).

Introducing the coordinate space wave functions for $\psi_A$ and
$\psi_{A-1}$ as:
\begin{equation} 
\psi_j(p_1,p_2,...p_j) = \left({1\over \sqrt{(2\pi)^3}}\right)^j 
 \int d^3x_1 d^3x_2,...d^3x_j 
e^{-i(\vec x_1\cdot \vec p_1+\vec x_2\cdot \vec p_2+...+\vec x_j\cdot\vec p_j)}
\phi_A(x_1,x_2,...x_j),
\label{wf_x}
\end{equation}
where $j\equiv A,A-1$ we can represent the IA amplitude as follows:
\begin{eqnarray}
T^{(a)}   & = &  
\int d^3x_1d^3x_2d^3x_3
\phi_A(x_1,x_2,x_3)F^{em}_1(Q^2) 
e^{i\vec x_1\cdot\vec q}\phi^{\dag}_{A-1}(x_2,x_3) e^{-i\vec x_1\cdot\vec p_f} = 
 \nonumber \\
& & 
\int d^3x_1d^3x_2d^3x_3
\phi_A(x_1,x_2,x_3)F^{em}_1(Q^2)
e^{-i{3\over 2}\vec p_m\cdot\vec x_1}\phi^{\dag}_{A-1}(x_2-x_3).
\label{amp_0_3}
\end{eqnarray}
We define $x_1$, as the  coordinate of the struck nucleon,  
and $x_2$, $x_3$ as  coordinates of the  residual nuclear system. 
In the last part of eq.(\ref{amp_0_3}) we introduce the wave function 
of residual nuclear system with separated internal and center of mass 
(CM) motion by:
\begin{equation}
\phi_{A-1}(y_2,y_3) = \phi(y_{2}-y_{3})e^{i{\vec y_{2}+\vec y_{3}\over 2}
\cdot\vec p_{cm}},
\label{wf_cmi}
\end{equation}
with $\vec p_{cm}$ is CM momentum of residual two-nucleon system.

Obviously, as follows from eqs.(\ref{amp_0_2},\ref{amp_0_3}),
within the impulse approximation measuring  $q$ and  $p_f$  
one directly measures the Fermi momentum of a nucleon in the  nucleus:
$\vec p_1=\vec p_m = \vec p_f - \vec q$.

\section{Single rescattering amplitude}

The diagrams of  Fig.2b  and  Fig.2c describe the processes where the
struck (fast) nucleon rescatters   off one of  the spectator nucleons. 
The  general  expression for the amplitude corresponding to the   
diagram Fig.2b  is given by  $n=1$ term of eq.(\ref{amp_n}) as:
\begin{equation}  
T^{(b)} =  
\int {\Gamma(p_1,p_2,p_3)\over D(p_1) D(p_2) D(p_3)}F^{em}_1(Q^2)
{f^{NN}(p'_2-p_2)\over D(p_1+q)} {\Gamma(p'_2,p_3)\over  D(p'_2)} 
{d^4p_2\over i (2\pi)^4} {d^4p_3 \over i (2\pi)^4}
\label{eq.6}
\end{equation}
where
\begin{equation}
p_1 = {\cal P}_A -p_2 - p_3; \ \ \ \ \ \ \ \ \ \  p'_2 = {\cal P}_{A-1}- p_3.
\label{eq.7}
\end{equation}
Our  interest is in the kinematics where   contribution of  the 
vacuum diagram is negligible, thus as in the previous section,  
we can perform the integration over the $d^0p_2d^0p_3$ by taking   
residues over the poles in the nucleon propagators  $D(p_2)^{-1}$ 
and   $D(p_3)^{-1}$. The integration  results in the  replacement 
$$\int {d p^0_{2,3}\over 2\pi i} {1\over D(p_{2,3})}\rightarrow 
{1\over 2 E_{2,3}}\approx {1\over 2m}.$$  
After the integrations over $d^0p_2d^0p_3$ are performed, the diagram 
of Fig.2b becomes the noncovariant time ordered  diagram, where a 
virtual photon is  absorbed by the target nucleon, and then  
the  produced fast nucleon rescatters off a  spectator nucleon:
\begin{equation}
T^{(b)}  = 
{1\over \sqrt{2m}\cdot (2m)^2}
\int 
{\Gamma(p_1,p_2,p_3)\over D(p_1)}F^{em}_1(Q^2)
{f^{NN}(p'_2-p_2)\over D(p_1+q)} {\Gamma(p'_2,p_3)\over  D(p'_2)} 
{d^3p_2\over  (2\pi)^3} {d^3p_3 \over (2\pi)^3}
\label{eq.6b}
\end{equation}
The definition of the momentum space wave functions is now
straightforward. It corresponds to the nonrelativistic reduction 
of the nuclear vertices $\Gamma_A$ and $\Gamma_{A-1}$ as given by 
eq.(\ref{wf_p}). Hence we obtain:
\begin{equation}
T^{(b)} =  {\sqrt{(2\pi)^3}(2\pi)^3 \over  2m } 
\int \psi_A(p_1,p_2,p_3)F^{em}_1(Q^2){f^{NN}(p'_2-p_2)\over D(p_1+q)} 
\psi_{A-1}(p'_2,p_3){d^3p_1 \over (2\pi)^3 } {d^3p_3\over (2\pi)^3}.
\label{eq.9}
\end{equation}

Here $D(p_1+q)^{-1}$  describes the struck nucleon propagator  
before the rescattering:
\begin{eqnarray} 
- D(p_1+q) & = & (p_1+q)^2 - m^2 + i\epsilon =  
p_1^2 + 2p_1q + q^2 - m^2 +i\epsilon \approx   \nonumber \\ 
& \approx & 2q\left[{2mq_0-Q^2\over 2q} - p_{1z} + i\epsilon\right],
\label{eq.10}
\end{eqnarray}
where $q \equiv \left|\vec q \right|$ and because of nonrelativistic  
approximation for nuclear wave function (see Appendix A)
we neglect  ${p_{1}^2\over 2m^2}$ as compared to $1$.
The factor ${2mq_0-Q^2\over 2q}$ is fixed by the external kinematics, 
since both $q_0$ and $Q^2$ are measured. It follows from 
$(q+p_A-p_{A-1})^2=m^2$ that:
\begin{equation}
{2mq_0-Q^2\over 2q} = p^{m}_z + {q_0\over q}(m + E_{A-1}-M_A)
+ {m^2-\tilde m^2\over 2 q}\approx p^{m}_z + \Delta_0,
\label{eq.11}
\end{equation} 
where $p^{m}_z = p_{fz}-q$, $\tilde m^2 \equiv [p_A-p_{A-1}]^2$ is the
virtuality of the interacting nucleon,  and
\begin{equation}
\Delta_0= {q_0\over q} (m + E_{A-1}-M_A)\equiv {q_0\over q}E_{m},
\label{eq.Delta}
\end{equation}
where $E_m = q_0 +m - \sqrt{m^2+p_f^2}$ is the missing energy in the 
reaction. In  the case of the  three body breakup kinematics for  
the scattering off the  $^3He$ target $E_m =  T_{pn}+|\epsilon_b|$, where 
$T_{pn}$ is  the kinetic energy of the  spectator (two nucleon) system 
and $|\epsilon_b|$ - is the modulus of the target binding energy. 
In the right-hand  side of eq.(\ref{eq.11})  we neglected 
the term ${m^2-\tilde m^2\over 2 q}$ related to the virtuality of 
interacting nucleon since at fixed recoil energy this factor 
is of the order of $O(E_m/q)$ and its contribution decreases with 
increase of the transferred momentum  $q$.

Inserting eq.(\ref{eq.11}) into the expression for the propagator of 
knocked-out nucleon(\ref{eq.10}) and redefining  the  $NN$ scattering 
amplitude as $f^{NN}/2qm= f^{NN}$, to be in accordance with the optical 
theorem in the form: $Im f^{NN}(t=0) = \sigma^{NN}_{tot}$,  we obtain:
\begin{eqnarray} 
T^{(b)} = -{\sqrt{(2\pi)^3}(2\pi)^3 \over  2 } 
\int \psi_A(p_1,p_2,p_3)F^{em}_1(Q^2){f^{NN}(p'_2-p_2)\over [p^{m}_z + \Delta_0 
- p_{1z} + i\epsilon]} \psi_{A-1}(p'_2,p_3) \nonumber \\
{d^3p_1\over (2\pi)^3} {d^3p_3\over (2\pi)^3}.
\label{eq.12}
\end{eqnarray}

We can perform   integration over $p_{1z}$ by transforming integrals 
into the coordinate space representation  and using the fact that 
for the soft $NN$ scatterings,  at high energies, 
$f^{NN}(p'_2-p_2)\approx f^{NN}(p'_{2t}-p_{2t})$. 
Using the coordinate space representation of the nuclear wave 
functions given by eq.(\ref{wf_x}) and the coordinate space  
representation of the nucleon propagators
\begin{equation}
{1\over [p^{m}_z + \Delta_0 
- p_{1z} + i\epsilon]} = 
-i\int\Theta(z^0)e^{i(p^z_{m}+\Delta_0-p_{1z})z^0}dz^0,
\label{eq.18}
\end{equation}
we obtain for the single rescattering amplitude $F^{(b)}$ 
(see Appendix B):
\begin{eqnarray}
T^{(b)} & = & {i\over  2} 
\int  d^3x_1 d^3x_2 d^3x_3 \phi_A(x_1,x_2,x_3)
\Theta(z_2-z_1)e^{i(\vec b_2-\vec b_1)\cdot\vec k_t}F^{em}_1(Q^2)
\nonumber  \\
& & f^{NN}(p'_{2t}-p_{2t}) e^{i\Delta_0(z_2-z_1)}
e^{-i{3\over 2}\vec p_{m}\cdot\vec x_1}\phi^{\dag}(x_2-x_3){d^2k\over (2\pi)^2},
\label{eq.19}
\end{eqnarray}
where  $\vec k_t = \vec p^{\ t}_{1}-\vec p^{\ t}_{m}= \vec p{\ '}_{2t}-\vec p_{2t}$ is the 
momentum transferred in the rescattering and $\vec b_{1}$, $\vec b_{2}$ 
are transverse  components of the vectors 
$\vec x_1$, and $\vec x_2$. It is convenient to introduce the  
generalized profile function\cite{FGMSS95}:
\begin{equation}
\Gamma^{NN}(x,\Delta) =  {1\over 2i}e^{i\Delta z} 
\int f^{NN}(k_t)e^{i\vec b\cdot \vec k_t}{d^2k_t\over (2\pi)^2}.
\label{eq.20}
\end{equation}
Using $\Gamma^{NN}(x,\Delta)$, we can write eq.(\ref{eq.19}) in a form resembling 
the Glauber theory expression for single rescattering:
\begin{eqnarray}
T^{(b)} & =  & -  \int  d^3x_1 d^3x_2 d^3x_3 
\phi_A(x_1,x_2,x_3)F^{em}_1(Q^2)\Theta(z_2-z_1)\Gamma^{NN}(x_2-x_1,\Delta_0) 
\nonumber  \\
& & \ \ \ \ \ \ e^{-i{3\over 2}\vec p_{m}\cdot\vec x_1}\phi^{\dag}(x_2-x_3).
\label{eq.19n}
\end{eqnarray}
The amplitude for the single rescattering - $T^{(c)}$, corresponding
to Fig.2c, can be obtained from eq.(\ref{eq.19n}) replacing 
$r_2\leftrightarrow r_3$. Thus the whole amplitude, which includes 
IA and the  single rescattering contributions is: 
\begin{eqnarray}
T^{(a)} + T^{(b)}+T^{(c)}  =  \int  d^3x_1 d^3x_2 d^3x_3 
\phi_A(x_1,x_2,x_3)F^{em}_1(Q^2) 
\left[1+ \hat T^{(1)}_{FSI}\right]\times  \nonumber \\
e^{-i{3\over 2}\vec p_{m}\cdot\vec x_1}\phi^{\dag}(x_2-x_3),
\label{eq.19u}
\end{eqnarray}
where 
\begin{equation}
\hat T^{(1)}_{FSI} =\Theta(z_2-z_1)\Gamma^{NN}(x_2-x_1,\Delta_0) +
\Theta(z_3-z_1)\Gamma^{NN}(x_3-x_1,\Delta_0)
\label{single_resc} 
\end{equation}
is the operator of FSI corresponding to the single rescattering 
contribution. Eq.(\ref{single_resc}) can be generalized for  the 
scattering off a nucleus with atomic number $A$  as follows:
\begin{equation}
\hat T^{FSI(1)} = 
1 + \sum\limits_{j=2}^{A} \Theta(z_j-z_1)\Gamma^{NN}(x_1-x_j,\Delta_0).
\label{eq.19g}
\end{equation}
The deduced operator for FSI have the form analogous to the familiar
operator deduced within nonrelativistic Glauber approximation 
\cite{Glauber}. The key difference is that the profile function - 
$\Gamma$  is  modified by  the additional phase factor $e^{i\Delta z}$.  
This factor accounts for the geometry of high-energy processes 
related to the longitudinal momentum transfer in the rescattering. 
Note that similar  factor is present in the expressions for the  
diffractive photoproduction of vector mesons\cite{Yennie,BSYP}, where 
it accounts for the difference between the masses of final vector 
mesons and initial (virtual) photon ($|t_{min}|>0$). In this case it 
reflects finite longitudinal distances ( $\le R_A$) for photoproduction 
at intermediate energies. In our case the factor  $\Delta_0$ arises 
from excitations in the residual nuclear system (see eq.(\ref{eq.Delta})). 

\begin{figure}[h]
\centerline{
\psfig{file=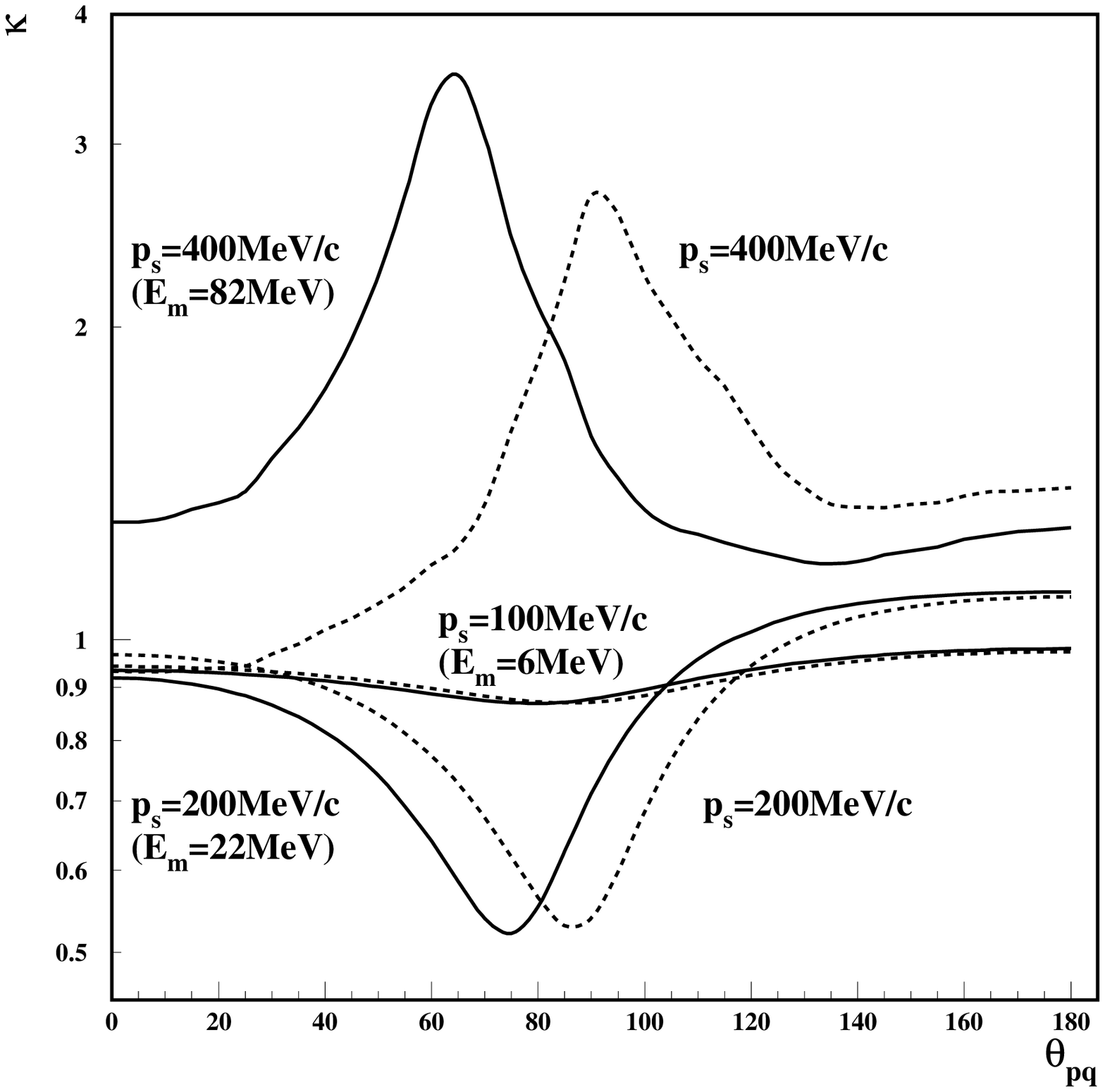,width=14cm,height=10cm}
}
\end{figure}

\vspace{0.3cm}

\noindent {\bf Fig. 3} 
\ {\em  
Dependence of the ratio $\kappa$, of the $d(e,e'p)n$  cross section 
calculated including the IA and FSI terms to the cross section 
which includes IA term only, on the angle 
$\Theta_{\vec p_s \vec q}\equiv \hat{\vec p_s \vec q}$ for different 
spectator momenta $p_s$. 
Solid line corresponds to FSI, calculated according 
to eq.(\ref{eq.12}-\ref{eq.19g}), dashed line corresponds to FSI 
calculated according to conventional Glauber approximation. }

\vspace{0.4cm}

To illustrate  the importance of derived modification of scattering 
operator we calculate the cross section of  $(e,e'N)$ scattering 
off the deuteron target where eq.({\ref{eq.19g}) provides the complete 
form for the FSI operator\cite{FGMSS95}. Fig.3 represents the ratio 
of the full cross section to the cross section  calculated within  
the impulse approximation. For  small momenta of a target nucleon 
($p_s< 100~MeV/c$) or for small excitation energies 
($E_{m}\approx {p_s^2\over 2m}$) 
predictions of generalized eikonal approximation (solid line)  
and conventional Glauber approximation (dashed line) coincidences.  
This demonstrates the consistency of our approach with Glauber  
approximation where target nucleons are interpreted  as a stationary 
scatterers and their Fermi momenta has been neglected. However at larger 
Fermi momenta (or excitation energies) predictions of both approaches 
are considerably different. For example for $p_s=400~MeV/c$ the
prediction for  angular dependence of the maximal contribution from 
the rescattering amplitude (i.e. the position of the maximum in Fig.3) 
differs as much as by $30^0$. 
Such a difference is quite dramatic and can be checked in the 
forthcoming experiments at Jefferson Laboratory\cite{KG,EGM}.
Practically the same difference arises for example in the break up of
$^3He$ if one of spectators has momentum $p \sim 0$\cite{EGM}.

\section{Double rescattering amplitude}

The diagrams of Fig.2d,2e describe the amplitude of the process
where struck nucleon rescatters sequentially off both spectator nucleons. 
From eq.(\ref{amp_n}) choosing $n=2$, for the double rescattering amplitude 
of Fig.2d, we obtain:
\begin{eqnarray}  
& & T^{(d)}   =  \nonumber  \\
& & \int 
{\Gamma(p_1,p_2,p_3)\over D(p_1) D(p_2) D(p_3)} F^{em}_1(Q^2)
{f^{NN}(p'_2-p_2)\over D(p_1+q)} {f^{NN}(p'_3-p_3)\over D(p_1+q+p_2-p'_2)} 
\nonumber \\
& & {\Gamma(p'_2,p'_3)\over  D(p'_2) D(p'_3)} \delta^4(p_A- p_2 - p_3 - p_1) 
 \delta^4(p_{A-1}- p'_2 - p'_3) d^4p_1 d^4p_2 d^4p_3  d^4p'_2 d^4p'_3 
\left[{1\over i (2\pi)^4}\right]^3
=  \nonumber \\ 
& & \int 
{\Gamma(p_1,p_2,p_3)\over D(p_1) D(p_2) D(p_3)}F^{em}_1(Q^2)
{f^{NN}(p'_2-p_2)\over D(p_1+q)}{f^{NN}(p'_3-p_3)\over D(p_1+q+p_2-p'_2)} 
\times \nonumber \\
& & \ \ \ \ \ \ \ \ {\Gamma(p'_2,p'_3)\over  D(p'_2)D(p'_3)} 
{d^4p_2\over i(2\pi)^4} {d^4p_3\over i(2\pi)^4}{d^4p'_3 \over i(2\pi)^4},
 \label{eq.25}
\end{eqnarray}
where
\begin{equation}
p_1 = {\cal P}_{A} - p_3 - p_2; \ \ \ \ \ \ \ 
p'_2 = {\cal P}_{A-1} - p'_3.
\label{eq.26}
\end{equation}
Then, using the same approximations as for the cases of IA and single 
rescattering amplitudes we can perform integration over $d^0p_2$, $d^0p_3$, 
$d^0p'_3$, which effectively results in the replacement 
$\int{d^0p_{j}\over 2\pi i D(p_{j})} 
\rightarrow  {1\over 2 E_{j}} \approx {1\over 2m}$, ($j=2,3,3'$).

Using  eq.(\ref{eq.26}) and the definition of the initial and final state 
wave functions from Section~\ref{IA} we obtain:
\begin{eqnarray}
T^{(d)} & = &  {\sqrt{(2\pi)^3}(2\pi)^3\over 4m^2} 
\int \psi_A(p_1,p_2,p_3)F^{em}_1(Q^2){f^{NN}(p'_2-p_2)\over D(p_1+q)}
{f^{NN}(p'_3-p_3)\over D(p_1+q+p_2-p'_2)}\times \nonumber \\ 
&  &  \psi_{A-1}(p'_2,p'_3)
{d^3p_1\over (2\pi)^3}{d^3p_3\over (2\pi)^3}{d^3p'_3\over (2\pi)^3},
\label{eq.27}
\end{eqnarray}
where $D(p_1+q)$ is given by  eq.(\ref{eq.10}). Using eq.(\ref{eq.26})
we can rewrite $D(p_1+q+p_2-p'_2)$ as: 
\begin{eqnarray}
- D(p_1+q+p_2-p'_2) & = & - D(q+p_{A}-p_{A-1}+p'_3-p_3) \nonumber \\ 
& = &  (q+p_{A}-p_{A-1}+p'_3-p_3)^2-m^2 + i\epsilon  \approx \nonumber \\
& &   2q\left[{q_0\over q}(E'_3-E_3)-(p'_{3z}-p_{3z})+i\epsilon\right]
= \left[(\Delta_{3} - (p'_{3z}-p_{3z}) + i\epsilon\right]. \nonumber \\
\label{eq.28}
\end{eqnarray}
In the derivation of eq.(\ref{eq.28}) we use the kinematic condition
for the quasielastic scattering: $(q+p_A-p_{A-1})^2=m^2$ and define 
$\Delta_{3} = {q_0\over q}(E'_3-E_3)$. Similar to  the previous 
section after redefining  the $NN$ amplitude as $
f^{NN}/2qm\rightarrow f^{NN}$ we obtain: 
\begin{eqnarray}
T^{(d)} & = & {\sqrt{(2\pi)^3}(2\pi)^3\over 4} 
\int \psi_A(p_1,p_2,p_3)F^{em}_1(Q^2)\times \nonumber  \\ 
&  & {f^{NN}(p'_2-p_2)\over p^{m}_z+\Delta_0-p_{1z}+i\epsilon}
\cdot {f^{NN}(p'_3-p_3)\over \Delta_{3} - (p'_{3z}-p_{3z})+i\epsilon} 
\psi_{A-1}(p'_2,p'_3)
{d^3p_1\over (2\pi)^3}{d^3p_3\over (2\pi)^3}{d^3p'_3\over (2\pi)^3}.
\label{eq.29}
\end{eqnarray}
Integration in eq.(\ref{eq.29}) can be performed in the  coordinate 
space,  using the Fourier transformation of the wave functions 
according to eq.(\ref{wf_x}) and nucleon propagators according to  
eq.(\ref{eq.18}). For the  double rescattering amplitude, $T^{(b)}$, 
we obtain (see Appendix B):
\begin{eqnarray} 
T^{(d)}
& = & {i^2\over  4} 
\int d^3x_1 d^3x_2 d^3x_3 
\phi_A(x_1,x_2,x_3) F^{em}_1(Q^2) \nonumber \\ 
& \times & \left [\Theta(z_2-z_1)f^{NN}(k_{2t})
e^{i\vec k_{2t}\cdot(\vec b_2-\vec b_1)}
e^{i(\Delta_0-\Delta_{3})(z_2-z_1)}{d^2k_2\over (2\pi)^2}\right]\nonumber \\ 
& \times & \left [\Theta(z_3-z_2)f^{NN}(k_{3t})
e^{i\vec k_{3t}\cdot (\vec b_3-\vec b_1)}
e^{i\Delta_{3}(z_3-z_1)}{d^2k_3\over (2\pi)^2}\right]
e^{-i{3\over 2}\vec x_1\cdot \vec p_{m}} \phi^{\dag}(x_2-x_3),
\label{eq.33}
\end{eqnarray}
where $k_{2t}$ and $k_{3t}$ are the momenta transferred in the first 
and second rescattering vertices in Fig.2d.

For a complete calculation of the double rescattering term one should 
take into account the amplitude $F^{(e)}$ too, which corresponds to Fig.2e. 
This amplitude can be derived from eq.(\ref{eq.33}) by interchanging  
coordinates of nucleons "2" and "3".  Finally, using definition of the 
modified profile functions from eq.(\ref{eq.20}) we obtain for 
$T^{(d)} + T^{(e)}$:
\begin{eqnarray}
\hat T^{(2)}\equiv T^{(d)} + T^{(e)} 
& = & \int d^3x_1 d^3x_2 d^3x_3 \phi_A(x_1,x_2,x_3) 
F^{em}_1(Q^2) 
{\cal O}^{(2)}(z_1,z_2,z_3,\Delta_0,\Delta_{2},\Delta_{3})\nonumber \\
& & \Gamma^{NN}(x_2-x_1,\Delta_0)\Gamma^{NN}(x_3-x_1,\Delta_0)
e^{-i{3\over 2}\vec x_1 \cdot\vec p_{m}} \phi^{\dag}(x_2-x_3).
\label{eq.35}
\end{eqnarray}
Here $\hat T^{(2)}_{FSI}$ - is the operator of FSI describing
the double rescattering contribution and we introduce the 
${\cal O}$ function which accounts for the geometry of two 
sequential rescatterings as:
\begin{eqnarray} 
{\cal O}^{(2)}(z_1,z_2,z_3,\Delta_0,\Delta_{2},\Delta_{3}) & = &  
\nonumber \\ 
& &  \Theta(z_2-z_1)\Theta(z_3-z_2)e^{-i\Delta_{3}(z_2-z_1)}
e^{i(\Delta_{3}-\Delta_0)(z_3-z_1)} \nonumber \\ 
& & + \Theta(z_3-z_1)\Theta(z_2-z_3)e^{-i\Delta_{2}(z_3-z_1)}
e^{i(\Delta_{2}-\Delta_0)(z_2-z_1)}.
\label{eq.36}
\end{eqnarray}

Eqs.(\ref{amp_0_3},\ref{eq.19n},\ref{eq.19u},\ref{eq.35}) represent 
the complete set of scattering amplitudes necessary to calculate 
knock-out reactions off the $^3He(^3H)$ target:
\begin{eqnarray}
T^{(a)} + T^{(b)}+T^{(c)}+T^{(d)}+T^{(e)}  = 
 \int  d^3x_1 d^3x_2 d^3x_3 \phi_A(x_1,x_2,x_3)F^{em}_1(Q^2)
\nonumber \\
 \left[1+ \hat T^{(1)}_{FSI} +\hat T^{(2)}_{FSI} \right]
e^{-i{3\over 2}\vec p_{m}\cdot\vec x_1}\phi^{\dag}(x_2-x_3).
\label{amp_sum3}
\end{eqnarray}
It is worth to note that in the derivation of above formulae no 
specific assumptions have been made  on the  nuclear wave functions. 
Therefore realistic wave functions of nuclei can be implemented 
to calculate the high-energy knock-out reactions for different 
configurations of the  residual two-nucleon system.

Eq.(\ref{eq.35}) can be generalized to calculate the double rescattering 
amplitude for $(e,e'N)$ reactions off $A$ nucleus as follows:
\begin{equation}
\hat T_{FSI}^{(2)} = \sum\limits_
{i,j=2;i\ne j}^{A} 
{\cal O}^{(2)}(z_1,z_i,z_j,\Delta_0,\Delta_{i},\Delta_{j}) 
\Gamma^{NN}(x_i-x_1,\Delta_0)\Gamma^{NN}(x_j-x_1,\Delta_0).
\label{eq.37}
\end{equation}
Generalization of the FSI operator $ \hat T_{FSI}^{(2)}$ to multiple 
rescatterings is straightforward: 
\begin{eqnarray}
\hat T_{FSI}^{(n)} = \sum\limits_
{i,j, ..n=2;i\ne j\ne .. n}^{A} 
{\cal O}^{(n)}(z_1,z_i,z_j,...z_n,
\Delta_0,\Delta_{i},\Delta_{j}...\Delta_{n}) \times
\nonumber \\
\Gamma^{NN}(x_i-x_1,\Delta_0)\cdot\Gamma^{NN}(x_j-x_1,\Delta_0)\cdot...\cdot
\Gamma^{NN}(x_n-x_1,\Delta_0),
\label{eq.38}
\end{eqnarray}
where 
\begin{eqnarray}
{\cal O}^{(n)}(z_1,z_i,z_j,...,z_n\Delta_0,
\Delta_{i},\Delta_{j}...\Delta_{n}) = 
\sum\limits_{perm} \Theta(z_i-z_1)\Theta(z_j-z_i)...\Theta(z_n-z_{n-1})\times
\nonumber \\ 
e^{i(\Delta_0 - \Delta_j-...\Delta_n)(z_i-z_1)} 
e^{i\Delta_j(z_j-z_1)}...e^{i\Delta_n(z_n-z_1)}
e^{-i\Delta_0(z_i+z_j+...z_n - n\times z_1)}.
\label{eq.39}
\end{eqnarray}
The sum in eq.(\ref{eq.39}) goes over all permutations between $i,j,...n$. 
We would like to draw attention that the contribution of diagrams where 
ejected nucleon interacts with say nucleon "2" then with nucleon "3" 
and then again with nucleon "2" is exactly zero. In coordinate 
representation this follows  from the structure of the product of 
$\Theta$-functions. In the  momentum representation this follows from 
the possibility to close the contour of integration in the complex plane 
without encountering nucleon poles (see discussion in Section II.C of
Ref.\cite{FPSS}).

It is easy to check that  in the case of small excitation energies i.e. 
($\Delta_0$, $\Delta_{i}$,$\Delta_{j}$ ... $\Delta_{n}\rightarrow 0$):
\begin{equation} 
 {\cal O}^{(n)}(z_1,z_i,z_j,..z_n,\Delta_0,\Delta_{i},\Delta_{j},...
\Delta_n)
\mid_{\Delta_0, \Delta_{i,k,n}\rightarrow 0 } 
\Rightarrow \Theta(z_i-z_1)\Theta(z_j-z_1)...\Theta(z_n-z_1),
\label{eq.40}
\end{equation}
and eqs.(\ref{eq.37},\ref{eq.38}) are reduced  to the conventional form of 
the Glauber approximation, with a simple product of the $\Theta$-functions.
Within this particular approximation the sum over all $n$-fold rescattering
amplitudes can be represented in the form of optical potential.

However, usually in high-energy $(e,e'N)$ reaction the excitation energies 
are not too small. The use of the 
${\cal O}^{(n)}(z_1,z_i,z_j,...z_n,\Delta_0,\Delta_{i},\Delta_{j},...\Delta_n)$, 
defined according to eq.(\ref{eq.39}) instead of  simple product 
of $\Theta$ functions is the generalization of nonrelativistic Glauber
approximation to the processes where comparatively large excitation
energies are important.
The practical consequence of the difference between ${\cal O}^{(n)}$
and usual $\Theta$ functions is that for  sufficiently large 
excitation  energies the  sum of   $n$-fold rescatterings  
differs substantially from  the simple optical model limit.

To illustrate the deviations from the conventional Glauber 
approximation (which is expressed by using a  simple product 
of the $\Theta$ functions) in Fig.4  
we compare ${\cal O}^{(2)}(z_1,z_2,z_3,\Delta_0,\Delta_{1},\Delta_{2})$ 
function with $\Theta(z_2-z_1)\Theta(z_3-z_1)$ for $(e,e'p)$
scattering  off $^3He$ target. We use the kinematics for three body
breakup in the final state. Figure demonstrates a considerable  
deviation between  ${\cal O}^{(2)}$ and the product of $\Theta$-functions
already at  comparatively low excitation energies. For example, the   
real parts differ by more than  $20\%$ already for $\sim 60 ~MeV$, 
leading to comparable difference of the double rescattering amplitude 
calculated including effects of longitudinal momentum transfer.
The detailed numerical studies of these effects will be presented elsewhere.

\begin{figure}[h]
\centerline{
\psfig{file=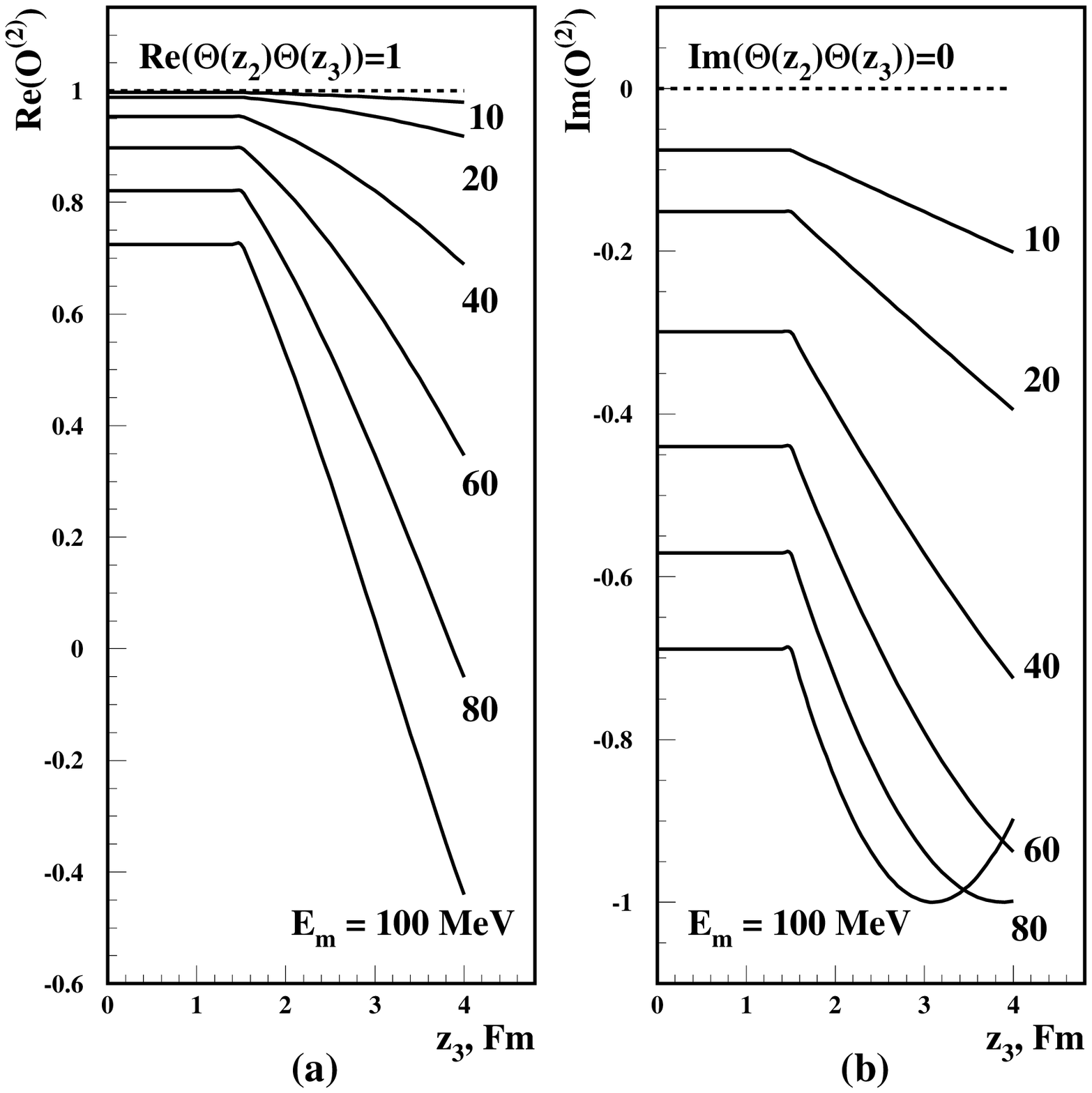,width=16cm,height=10cm}
}
\end{figure}

\vspace{0.3cm}

\noindent {{\bf Fig. 4} \ {\em  
Dependence of   ${\cal O}^{(2)}(z_1,z_2,z_3,\Delta_0,\Delta_{1},
\Delta_{2})$ and $\Theta(z_2-z_1)\Theta(z_3-z_1)$  on $z_3$ for  
different values of missing energy $E_m$ for $z_1=0$, $z_2=1.5~Fm$ 
and  $\Delta_{1}=\Delta_{2}=0$. 
a) Comparison of $Re {\cal O}^{(2)}(...)$(solid line) with  
$Re\Theta(z_2-z_1)\Theta(z_3-z_1)=1$(dashed line) and b) comparison  
of $Im {\cal O}^{(2)}(...)$(solid line) with 
$Im\Theta(z_2-z_1)\Theta(z_3-z_1)=0$ (dashed line).}}  
 
\vspace{0.4cm}

Thus we conclude, that the conventional Glauber approximation which 
neglects nuclear Fermi motion is applicable  in the case of small values 
of the residual nucleus excitation energies only.

 \section{FSI and the study of short-range nucleon correlations in nuclei}

It is generally believed that experimental condition 
$|\vec p_m|=|\vec p_f  - \vec q| > k_{F}$, (where $k_{F}\sim 250~MeV/c$ 
is momentum of Fermi surface for a given nucleus) will enhance 
the contribution to the cross section from  the short-range 
nucleon correlations in the nucleus wave function. However 
simple impulse approximation relation (eq.(\ref{amp_0_2})) is, 
in general, distorted by the FSI. Let us denote the internal 
momentum of the  knock-out nucleons prior to the collision 
as $\vec p_1(p_{1z},p_{1t})$. It follows from eqs.(\ref{eq.19}),(\ref{eq.33}):
\begin{equation}
\vec p_{1t} = \vec p_{mt} - \vec k_t,
\label{kt}
\end{equation}
where  $\vec p_{mt}$ the transverse component of the measured missing 
momentum, and $k_t$ is the  momentum transferred in rescattering.
Average $<\vec k_t^2 > \sim 0.1~ GeV^2$ in the integral over $k_t$  are 
determined by the slope of the $NN$ amplitude. The longitudinal 
component  of the nucleon momentum in the initial state can be 
evaluated through  its value in the pole of the rescattered 
nucleon propagator (see e.g. eqs.(\ref{eq.12}),(\ref{eq.29})):
\begin{equation}
p_{1z}    =   p^z_m + \Delta_0
\label{eq.pole}
\end{equation}
where $p^z_m$ is longitudinal component of the  measured missing momentum  
and $\Delta_0$  represents the excitation energy of the residual nuclear 
system (see eq.(\ref{eq.11})). $\Delta_0$ is always positive
(eq.(\ref{eq.Delta})). Thus, if  measured  $p_{zm} >k_{F}$  then, 
$p_{1z}$ is even larger, i.e. ($p_{z1} > p_{zm}$) and therefore the FSI 
amplitude is as sensitive to  the short range correlations as the 
IA amplitude. 
In particular, within the approximation when high-momentum component 
of the nuclear spectral function is due to two-nucleon short-range 
correlations \cite{FS81,CFSM} the condition $p_{zm}>k_{F}$ 
corresponds to projectile electron scattering  off the forward moving 
nucleon of the two-nucleon correlation accompanied by the emission of 
backward nucleon.

Situation is opposite if measured momentum $p_{zm} < - k_{F}$.
It follows from eq.(\ref{eq.pole}), that in this case the momenta in 
the wave function contributing to rescattering amplitude are smaller 
than those for IA: $|p^z_1| = |p^z_m|-\Delta_0  < |p^z_m|$. 

Experimentally, this situation corresponds to the forward nucleon 
electroproduction at   ${Q^2\over 2mq_0}\equiv x>1$. However an important  
feature in this case is that in exclusive electroproduction 
the value of $\Delta_0$ is measured  experimentally and can be easily chosen 
so  that  momenta entering in the  ground state wave function would be larger 
than  $k_{F}$. Therefore to investigate the short-range correlations in the  
$(e,e'p)$ reactions for   $x>1$, we have  to impose an additional condition: 
\begin{equation}
|p^z_m| -\Delta_0 > k_{F}
\label{addx}
\end{equation}
to suppress the contribution from large internucleon  distances. 

Overall, we observe that in order to  study  short-range 
nucleon correlations in $(e,e'N)$ reactions off nuclei  
with minimal distortions due the FSI effects it is 
advantageous to use  the $x<1$ kinematics especially
with detection of a backward going nucleon.

Above  results have a simple explanation in terms of the light-cone 
dynamics of  high-energy  scattering processes. Indeed, according to 
eq.(\ref{eq.Delta}) $\Delta_0$  does not disappear  with increase of energy. 
Hence the non-conservation of the longitudinal momentum of nucleons 
given by eq.(\ref{eq.pole}) : $p_{1z}-p^z_m=\Delta_0$ remains finite in the 
high-energy limit. However, the rescattering of an energetic knock-out
nucleon practically does not change the "-" component of its four-momentum 
$p_- \equiv E -p_z$, ($p_-$ is the longitudinal momentum as defined in 
light-cone variables, where $p^{\mu}\equiv p^{\mu}(p_+,p_-,p_t)$ with 
$p_{\pm}=E\pm p_{z}$).
Really, if we define $p_{1-} = m - p_{1z}$ and $p_{m-} = p_{f-} - q_- = 
m - E_{m} - p_{mz}$, where $E_m = m + E_{A-1}-M_A$  is the missing energy, 
then according to eq.(\ref{eq.pole}) the non-conservation of "-" component is:
\begin{equation}
p_{1-}-p_{m-} \approx {Q^2\over 2 q^2}E_m = {E_m\over 2(1+{q_0\over 2mx})}. 
\label{nonconv}
\end{equation}
It vanishes with increase of the  virtual photon energy $q_0$. Hence, 
the  physical interpretation of eq.(\ref{eq.pole}) is that at high energies 
elastic FSI does not change noticeably the  light-cone "-" component of 
the struck  nucleon momentum.  This reasoning indicates that description  
of the  FSI in high-energy processes should be simplified when treated
within the framework of the light-cone dynamics. Our previous analysis
of  $x>1$, large $Q^2$ data on inclusive $(e,e')$ processes 
is consistent with
this idea\cite{Day}. 

The above observation helps to rewrite the deduced formulae in the form  
accounting for, in the straightforward way, that high-energy processes 
develop along the light-cone.  

Let us introduce light-cone momenta $\alpha_i\equiv A{p_{i-}\over P_{A-}}$. 
Here $\alpha/A$ is a momentum fraction of target 
nucleus carried by the nucleon-$i$. Using the above discussed expressions for 
$p_{m-}$ and $p_{1-}$ and eqs.(\ref{eq.11},\ref{eq.Delta}) for the propagator 
of a  fast nucleon we obtain: 
\begin{equation}
{1\over [p^{m}_z + \Delta_0 
- p_{1z} + i\epsilon]} = {1\over m[\alpha_1 - \alpha_m +{q_0-q\over qm}E_m
+ i\epsilon]} \approx {1\over m[\alpha_1 - \alpha_m -{Q^2\over 2q^2}{E_m\over 
m}+ i\epsilon]}.
\label{lcprop}
\end{equation}
In the kinematics where relativistic effects in the wave function of the 
target and residual nucleus are small and $\alpha_j \approx 1-{p_{jz}\over m}$,  
there is a smooth correspondence between nonrelativistic and light-cone 
wave functions of the nucleus\cite{FS81}, 
i.e. $\phi_A(p_1,...p_j,..p_A)\approx \phi_A(\alpha_1,p_{1t}, ... 
\alpha_j,p_{jt},... \alpha_A,p_A)/m^{A\over 2}$. Therefore  the amplitude of single 
rescattering - eq.(\ref{eq.12}) can be rewritten as:
\begin{eqnarray} 
T^{(b)} = -{\sqrt{(2\pi)^3}(2\pi)^3 \over  2m } 
\int \psi_A(\alpha_1,p_{1t},\alpha_2,
p_{2t},\alpha_3,p_{3t})F^{em}_1(Q^2){f^{NN}\over 
[\alpha_1 - \alpha_m -{Q^2\over 2q^2}{E_m\over m}
+ i\epsilon]}\nonumber \\
 \psi_{A-1}(\alpha'_{2},p'_{2t},\alpha_3,p_{3t}) 
{d\alpha_1 d^2p_{1t}\over (2\pi)^3} {d\alpha_3 d^2p_{3t}\over (2\pi)^3}.
\label{eq.12lc}
\end{eqnarray}
where according to eq.(\ref{eq.7}) $\alpha_2 = \alpha'_2 = 3 -\alpha_1-\alpha_3$.  
Eq.(\ref{eq.12lc}) shows that in the limit when ${Q^2\over 2q^2}
{E_m\over m}\rightarrow 0$, the amplitude $T^{(b)}$ is expressed through  
the light-cone variables and the light-cone wave functions of nucleus. 
Note that   the eikonal scattering corresponds to the linear, in $\alpha_1$, 
propagator of the fast nucleon. It is instructive that regime of the 
light-cone  dynamics is reached in eq.(\ref{eq.12lc}) at relatively moderate 
energies. Indeed let as consider kinematics when $\alpha_1$ is close to unity, 
(which is the  case in our analysis). At $q_0\sim 2~GeV$, 
${Q^2\over 2q^2}{E_m\over m}= {1\over 2(1+{q_0\over 2mx})}{E_m\over m} 
\sim (0.05-0.07)\ll 1$. For estimate we take $x=1$ and for missing energy 
$E_m\sim 0.2-0.3~GeV$ which is close to the limit of applicability of 
the description of nuclei as a many-nucleon system (cf. \cite{FS88}).   
Similar reasoning is applicable for the double rescattering 
amplitude in eq.(\ref{eq.29}). Here we obtain:
\begin{eqnarray}
T^{(d)} & = & {\sqrt{(2\pi)^3}(2\pi)^3\over 4 m^2} 
\int
\psi_A(\alpha_1,p_{1t},\alpha_2,p_{2t},\alpha_3,p_{3t})F^{em}_1(Q^2)
\times \nonumber  \\ 
&  & {f^{NN}(p_{1t}-p_{mt}-(p'_{3t}-p_{3t}))\over 
[\alpha_1 - \alpha_m -{Q^2\over 2q^2}{E_m\over m} + i\epsilon]}
 {f^{NN}(p'_{3t}-p_{3t})\over 
[\alpha_3 - \alpha'_3 -{Q^2\over 2q^2}{k^2_{3t}\over2 m^2} + i\epsilon]}
\psi_{A-1}(\alpha_2,p'_{2t},\alpha_3,p'_{3t})
\\ \nonumber 
& & {d\alpha d^2p_{1t}\over (2\pi)^3}{d\alpha_3 d^2 p_{3t}\over (2\pi)^3}
{d\alpha'_3 d^2p'_{3t}\over (2\pi)^3}.
\label{eq.29lc}
\end{eqnarray}

Another interesting consequence of the  representation of the scattering 
amplitude through the light-cone variables, is the simple form of the
closure approximation for the  sum over the residual $(A-1)$ nuclear 
states in  $A(e,e'N)(A-1)$ reaction. 
When summing over  $E_m$ at fixed $p_m$ the rescattering amplitudes 
(cf. eq.(\ref{eq.12})) could not be factored out from the sum because they 
depend on $E_m$ through the $\Delta$ factors (cf. eq.(\ref{eq.Delta})).
In the case of the light-cone representation (cf. eq.(\ref{eq.12lc})) the 
analogous  procedure\cite{FS88} is to sum over   $p_+\approx m+E_m+p_{mz}$ 
at fixed $\alpha_m$. It  follows from  eqs.(\ref{eq.12lc}) and (\ref{eq.29lc}) 
that in such a sum the scattering amplitude is independent of  $p_+$  and 
therefore the application of closure has a simple form.

Note that the present discussion of the  light-cone dynamics is by no means 
complete, since we don't consider the relativistic effects which enter 
into the nuclear wave functions. The extension of the  current analysis 
to the light-cone formalism  will be presented elsewhere.

\section{Conclusions}

The Feynman diagram approach to the calculation of final state 
interactions at high energy $(e,e'N)$ reactions off nuclei provides 
a natural framework for  the generalization of the  conventional 
nonrelativistic  Glauber approximation to high-energy processes.  
This approach  adequately describes  also the light-cone dynamics 
characteristic at high-energy reactions.

It  follows from the consideration of Feynman diagrams that 
the formulae of the conventional Glauber approximation are a  
legitimate approximation for sufficiently small values of 
residual nuclear system's excitation energy (missing energy). 
Beyond this kinematic region, conventional approximation should be 
modified to describe correctly relativistic kinematics and the 
dynamics  of  FSI. This can  be done within the generalized 
eikonal  approach which is developed in the  paper for $(e,e'N)$ 
knock-out reactions.

The obtained formulae allow to find a kinematic domain preferable 
for the investigation of short-range nucleon correlations in nuclei.
We demonstrate that scattering off forward moving nucleons
(which corresponds to production of backward going nucleon spectators
from destruction of short-range pair correlations \cite{FS81}) is 
preferable for the investigation of  short-range nucleon correlations 
in nuclei.  We found  an additional kinematic condition: 
$|p^z_m|-\Delta_0 > k_{F}$ for semi-exclusive reactions to enhance the 
contribution of  short-range nucleon correlations at $x>1$
and reduce FSI.

We demonstrate also that dominance of  light-cone dynamics follows 
directly from the analysis of the Feynman diagrams, and that  the "-" 
component of the target nucleon momentum is almost conserved in FSI.
Therefore, by measuring the "-" component of the missing momenta we 
directly tag the preexisting momenta in the light cone nuclear wave 
function.

\begin{acknowledgements}
We would like to thank G.~A.~Miller for  useful suggestions and 
comments. The work was supported  in part by the Israel-USA Binational
Science Foundation Grant No.~92000126 and by the U.S. Department of Energy
under Contract No. DE-FG02-93ER40771.
\end{acknowledgements}

\vspace{0.4cm}

\appendix 
 
\vspace{0.4cm}
 
\section{Why closure approximation  in general is 
applicable in Light-Cone but not in the nucleus rest frame} 

In the calculation of $n$-fold rescattering amplitude of Fig.1 
we assumed the decoupling (from the excitation energies of 
intermediate states) of the propagator of high energy  knocked-out   
nucleon - $D(p_1+q)^{-1}$. Such a decoupling allows to  use the closure
over the sum over the excitations of intermediate nuclear states. 
As a result the scattering  amplitude in eq.(\ref{amp_n}), is calculated 
in terms of the propagators of free spectator nucleons in the 
intermediate states.

To visualize the conditions when the decoupling of high-energy 
part of the diagram of Fig.1 from low-energy part would be valid 
we consider two reference frame descriptions: Nucleus rest frame 
(Lab frame) and Light Cone.

In the Lab frame  the inverse propagator of energetic knocked-out  
nucleon: $-D(p_1+q)= (p_1+q)^2-m^2+i\epsilon$ can be written as:
\begin{equation}
(p_1+q)^2-m^2 = p_1^2 + 2E_1q_0 - 2\vec p_{1}\cdot{\vec q} + q^2 - m^2 = 
2|\vec q|\left[{p_1^2-m^2\over 2|\vec q|} + E_1{q_0\over |\vec q|} -
    p_{1z} - {Q^2\over |\vec 2q|}\right]
\label{a1}
\end{equation}
It follows from the right hand side of eq.(\ref{a1}) that only the term 
$E_1{q_0\over |\vec q|}- p_{1z}$ survives in limit of large momentum transfer 
($\vec q$) and fixed $x_{Bj}$. Thus in high energy  limit within Lab frame 
description one should retain the dependence of propagator on the 
excitation energy of intermediate state (via ${E_1}$). Therefore,
unless  the $E_1$ dependence of the propagator of knocked-out nucleon 
can be neglected the use of closure over the intermediate nuclear 
states can not be justified. In the Lab frame description such a 
neglection is legitimate in the nonrelativistic limit only where the 
term  ${p_1^2\over 2m^2}\ll 1$ is neglected everywhere in the expression 
of the scattering amplitude.
Such a restriction on the applicability of the closure for the sum over
the intermediate states is of crucial importance for the  models where 
relativistic effects are treated on the basis of the Lab frame description.

Above calculation does not take into account  additional approximate
conservation law characteristic for light-cone dynamics.
Let us introduce  light-cone momenta for  four-vectors as: 
$p^\mu(p_+,p_-,p_t)$, where $p_{\pm} = E\pm p_{z}$. Using these definitions, 
for the inverse propagator of knocked-out nucleon  one obtains the form:
\begin{equation}
(p_1+q)^2-m^2 = p_1^2 + p_{1+}q_{-} + p_{1-}q_{+} +q^2 - m^2 = 
q_{+}\left[{p_1^2-m^2\over q_+} + p_{1+}{q_{-}\over q_{+}} + 
p_{1-} - {Q^2\over q_{+}}\right].
\label{a2}
\end{equation}
As follows from the above equation that only term, that survives 
at fixed $x_{Bj}$ and high energy transfer limit, is $p_{1-}$. 
Therefore at fixed $p_{-}$ we found  effective factorization of  
high-energy propagator from low energy intermediate nuclear part 
whose excitation energy on light cone is defined  by the 
$p_{1+}$ \cite{FS81,FS88}.  Such a decoupling  applies for any 
values of Fermi momenta of  the target 
nucleon (no restriction like ${p_1^2\over 2m^2}\ll 1$ is needed). 
Therefore it is possible  to extend the applicability of the 
closure  over intermediate  states of the residual nucleus to 
the domain of relativistic  momenta of target nucleons.  The price 
is to introduce  the light-cone wave function's of the target (similar 
to the case of pQCD).

Note that the considerations in  present work are restricted by small 
Fermi momenta (eq.(\ref{kin1})) since  we use  ${p_1^2\over 2m^2}\ll 1$
in the scattering amplitude. For larger Fermi momenta a legitimate way 
to generalize obtained results is to use light-cone description, which
is out of scope of the present paper. Note that light-cone mechanics 
of nuclei is rather similar to the nonrelativistic ones\cite{FS81,FS91}.

\section {Coordinate space representation of the  scattering amplitude}

\subsection{ Single scattering amplitude}

We now will transform the single scattering amplitude of eq.(\ref{eq.12}) 
to coordinate space. Inserting the  configuration space representation 
of ground state and residual state wave functions according to 
eq.(\ref{wf_x}) into  eq.(\ref{eq.12}) and using energy-momentum 
conservation of eq.(\ref{eq.7}), for $T^{(b)}$ we obtain:
\begin{eqnarray} 
T^{(b)} & = & - {1\over 2} 
\int d^3x_1 d^3x_2 d^3x_3 d^3y_2 d^3y_3e^{-i(\vec x_1-\vec x_2)\cdot \vec p_1} 
\phi_A(x_1,x_2,x_3) {F^{em}_1(Q^2)f^{NN}(p'_{2t}-p_{2t})
\over [p^{m}_z + \Delta_0 - p_{1z} + i\epsilon]}  \nonumber  \\
&  & e^{i\vec y_2\cdot \vec p_{A-1}} 
e^{-i\vec p_3\cdot((\vec x_3-\vec x_2)-(\vec y_3-\vec y_2))}
\phi^{\dag}_{A-1}(y_2,y_3)
{d^3p_1\over (2\pi)^3} {d^3p_3 \over (2\pi)^3}  \nonumber \\ 
& = & - {1\over 2} 
\int d^3x_1 d^3x_2 d^3x_3 d^3y_2 d^3y_3e^{-i(\vec x_1-\vec x_2)\cdot \vec p_1} 
\phi_A(x_1,x_2,x_3){F^{em}_1(Q^2)f^{NN}(p'_{2t}-p_{2t})\over 
[p^{m}_z + \Delta_0 - p_{1z} + i\epsilon]} 
 \nonumber   \\
&  & e^{i\vec y_2\cdot \vec p_{A-1}} \delta^3((x_3-x_2)-(y_3-y_2))
\phi^{\dag}_{A-1}(y_2,y_3) {d^3p_1 \over (2\pi)^3}.
\label{eq.14}
\end{eqnarray}

Next, we introduce relative and CM coordinates as:
\begin{equation}
y_2 = {y_{23}\over 2} + y_{cm}; \ \ \ \ \ \ \ \ \ \ 
y_3 = {y_{23}\over 2} - y_{cm} ,
\label{eq.15}
\end{equation}
and separate   internal and CM motion of the recoil $pn$ system:
\begin{equation}
\phi_{A-1}(y_2,y_3) = \phi(y_{23})e^{iy_{cm}(p_p+p_n)}.
\label{eq.16}
\end{equation}
As a result,   $T^{(b)}$ takes the form: 
\begin{eqnarray} 
T^{(b)}  &  = &  - {1\over  2} 
\int d^3x_1 d^3x_2 d^3x_3 d^3y_2 d^3y_3e^{-i(\vec x_1-\vec x_2)\cdot\vec p_1} 
\phi_A(x_1,x_2,x_3){F^{em}_1(Q^2)f^{NN}(p'_{2t}-p_{2t})
\over [p^{m}_z + \Delta_0 - p_{1z} + i\epsilon]}  \nonumber \\ 
& & \ \ \ \ \ \ 
e^{-i{(\vec x_2-\vec x_3)\over 2}\cdot \vec p_{m}}\phi^{\dag}(x_2-x_3) 
{d^3p_1\over (2\pi)^3 }.
\label{eq.17a}
\end{eqnarray}
To integrate over $p_{1z}$ we  use the coordinate space representation
of the nucleon propagator ${1\over [p^{m}_z + \Delta_0 - p_{1z} +  i\epsilon]}$  
according to  eq.(\ref{eq.18}). Inserting  eq.(\ref{eq.18}) in  
eq.(\ref{eq.17a}), one can integrate  over  $p_{1z}$: $\int
\exp(-ip_{1z}(z_1-z_2+z^0))dp_{1z} =  2\pi\delta(z_1-z_2+z^0)$. 
After integrating over $dz^0$ we obtain:
\begin{eqnarray}
T^{(b)} & = & {i\over    2} 
\int  d^3x_1 d^3x_2 d^3x_3 
e^{i(\vec b_2-\vec b_1)\cdot(\vec p^{\/ t}_1-\vec p^{\/ t}_{m})}
\phi_A(x_1,x_2,x_3)F^{em}_1(Q^2) 
 \nonumber \\
& & f^{NN}(p'_{2t}-p_{2t})\Theta(z_2-z_1)
e^{-i{3\over 2}\vec p_{m}\cdot\vec x_1} e^{i\Delta_0(z_2-z_1)} \phi^{\dag}(x_2-x_3) 
{d^2p^t_1\over (2\pi)^2} 
 \nonumber \\
& = & {i\over  2} 
\int  d^3x_1 d^3x_2 d^3x_3 
\phi_A(x_1,x_2,x_3)
\Theta(z_2-z_1)e^{i(\vec b_2-\vec b_1)\cdot\vec k_t}F^{em}_1(Q^2)
 \nonumber  \\
& & f^{NN}(p'_{2t}-p_{2t})
 e^{i\Delta_0(z_2-z_1)}
e^{-i{3\over 2}\vec p_{m}\cdot\vec x_1}\phi^{\dag}(x_2-x_3){d^2k\over (2\pi)^2},
\label{eq.19ap}
\end{eqnarray}
where we define the momentum  transferred  in the rescattering 
as $\vec k_t = \vec p^{ \ t_1}-\vec p^{ \ t_{m}}= 
\vec p{\ '}_{2t}-\vec p_{2t}$ and $\vec b_{1}$, $\vec b_{2}$ are transverse  
components of vectors $\vec x_1$, $\vec x_2$.

\subsection{ Double scattering amplitude}

Integration in eq.(\ref{eq.29}) can be performed in the coordinate space,  
using the Fourier transform of the wave functions according to 
eq.(\ref{wf_x}{) and by introducing the $\vec L$ and $\vec k_3$:
\begin{equation}
\vec L = {\vec p_3\/' + \vec p_3 \over 2 }; \ \ \ \ \ \ \ \ 
\vec k_3  =\vec p_3\/' - \vec p_3.
\label{eq.30}
\end{equation}
Then for  $T^{(d)}$ we obtain:
\begin{eqnarray} 
T^{(d)} & = & {1\over   4} 
\int d^3x_1 d^3x_2 d^3x_3 d^3y_2 d^3y_3e^{-i(\vec x_1-\vec x_2)\cdot\vec p_1} 
\phi_A(x_1,x_2,x_3)F^{em}_1(Q^2)\nonumber \\ 
&  & {f^{NN}(p'_2-p_2)\over [p^{m}_z + \Delta_0 - p_{1z} + i\epsilon]}  
{f^{NN}(p'_3-p_3)\over [\Delta_{3} - (k_{3z})+i\epsilon]} \nonumber \\ 
&  & e^{i\vec y_2\cdot \vec p_{A-1}} e^{-i{\vec k_3\over 2}
\cdot[(\vec x_2-\vec x_3)+(\vec y_2-\vec y_3)]}
e^{i\vec L\cdot[(\vec x_2-\vec x_3)-(\vec y_2-\vec y_3)]}
\phi^{\dag}_{A-1}(y_2,y_3)
{d^3p_1\over (2\pi)^3}{d^3L\over (2\pi)^3}{d^3k_3\over (2\pi)^3}.
\label{eq.31}
\end{eqnarray}
Since we  consider  soft rescatterings of a high energy (knocked-out) 
nucleon off a slow spectator, we can use the observation
that the scattering  amplitude for two-body scattering - $f^{NN}(p'_3-p_3)$  
depends mainly on transverse components of transferred momentum $\vec k_{3t}$ 
and is practically independent of $\vec L$. Therefore we  can perform 
integration over $d^3L$ invoking the factor $\delta^3(x_2-x_3-(y_2-y_3))$. 
Similar to the previous section using  eqs.(\ref{eq.15}) and (\ref{eq.16})  
allows to  perform the integration  over  $d^3 y_{cm} d^3 y_{23}$:
\begin{eqnarray} 
T^{(d)} & = & {1\over   4} 
\int d^3x_1 d^3x_2 d^3x_3 e^{-i(\vec x_1-\vec x_2)\cdot\vec p_1} 
e^{-i\vec k_3\cdot(\vec x_2-\vec x_3)}
\phi_A(x_1,x_2,x_3)F^{em}_1(Q^2)  \nonumber \\
&  & {f^{NN}\over [p^{m}_z + \Delta_0 
- p_{1z} + i\epsilon]}  
{f^{NN}\over [\Delta_{3} - (k_{3z})+i\epsilon]} 
e^{-i{\vec x_2- \vec x_3\over 2} \vec p_{m}}\phi^{\dag}(x_2-x_3)
{d^3p_1\over (2\pi)^3} {d^3k_3\over (2\pi)^3}. 
\label{eq.32}
\end{eqnarray}
Furthermore we can take the integral over $p_{1z}$ similarly to 
the case of single rescattering amplitude, using the eq.(\ref{eq.18}). 
The integration by $k_{3z}$ can be  done using the   
representation: ${1\over \Delta_{3}-k_{3z}+i\epsilon} = - i\int\Theta(z^{k_3})
\exp{i(\Delta_{3}-k_{3z})z^{k_3}} d z^{k_3}$. 
The integration over $dp_{1z}$ and $dk_{3z}$ leads to the factor:
$2\pi\delta(z^0-(z_2-z_1))$ and  $2\pi\delta(z^{k_3}-(z_3-z_2))$  
 respectively.  After performing integration over $dz^0$ and  $dz^{k_3}$  
and  defining $\vec k_1 = \vec p^{\ t}_{1}-\vec p^{\ t}_{m}$ we obtain:
\begin{eqnarray} 
T^{(d)} & = & {i^2\over  4} 
\int d^3x_1 d^3x_2 d^3x_3 
\phi_A(x_1,x_2,x_3) F^{em}_1(Q^2)\nonumber \\ 
& \times & \left [\Theta(z_2-z_1)f^{NN}(k_{1t}-k_{3t})
e^{i\vec k_{1t}\cdot (\vec b_2-\vec b_1)}
e^{i\Delta_0(z_2-z_1)}{d^2k_1\over (2\pi)^2}\right] \nonumber \\ 
& \times & \left [\Theta(z_3-z_2)f^{NN}(k_{3t})
e^{i\vec k_{3t}\cdot(\vec b_3-\vec b_2)}
e^{i\Delta_{3}(z_3-z_2)}{d^2k_3\over (2\pi)^2}\right] 
e^{-i{3\over 2}\vec x_1 \cdot\vec p_{m}} \phi^{\dag}(x_2-x_3) \nonumber \\ 
& = & {i^2\over  4} 
\int d^3x_1 d^3x_2 d^3x_3 
\phi_A(x_1,x_2,x_3) F^{em}_1(Q^2) \nonumber \\ 
& \times & \left [\Theta(z_2-z_1)f^{NN}(k_{2t})
e^{i\vec k_{2t}\cdot(\vec b_2-\vec b_1)}
e^{i(\Delta_0-\Delta_{3})(z_2-z_1)}{d^2k_2\over (2\pi)^2}\right]\nonumber \\ 
& \times & \left [\Theta(z_3-z_2)f^{NN}(k_{3t})
e^{i\vec k_{3t}\cdot(\vec b_3-\vec b_1)}
e^{i\Delta_{3}(z_3-z_1)}{d^2k_3\over (2\pi)^2}\right]
e^{-i{3\over 2}\vec x_1 \cdot\vec p_{m}} \phi^{\dag}(x_2-x_3),
\label{eq.33ap}
\end{eqnarray}
where at the last step we do the replacement 
$\vec k_{1t}\rightarrow \vec k_{2t}+ \vec k_{3t}$.

\end{document}

%% file: figure2.tex
\bigphotons

\begin{picture}(20000,10000)(500,0)
\THICKLINES
\drawline\fermion[\E\REG](0,-100)[3000]
\drawline\fermion[\E\REG](0,-50)[3000]
\drawline\fermion[\E\REG](0,00)[3000]
\drawline\fermion[\E\REG](0,50)[3000]
\drawline\fermion[\E\REG](0,100)[3000]
\put(4000,150){\circle{2000}}
\drawline\fermion[\E\REG](0,150)[3000]
\drawline\fermion[\E\REG](0,200)[3000]
\drawline\fermion[\E\REG](0,250)[3000]
\drawline\fermion[\E\REG](0,300)[3000]
\drawline\fermion[\E\REG](0,350)[3000]
\drawline\fermion[\E\REG](0,400)[3000]
\drawline\fermion[\E\REG](4700,900)[13200]
\drawline\scalar[\E\REG](5000,-200)[6]
\drawline\fermion[\E\REG](4300,-850)[14000]
\put(18500,150){\circle{2000}}
\drawline\fermion[\E\REG](19500,-100)[3000]
\drawline\fermion[\E\REG](19500,-50)[3000]
\drawline\fermion[\E\REG](19500,00)[3000]
\drawline\fermion[\E\REG](19500,50)[3000]
\drawline\fermion[\E\REG](19500,100)[3000]
\drawline\fermion[\E\REG](19500,150)[3000]
\drawline\fermion[\E\REG](19500,200)[3000]
\drawline\fermion[\E\REG](19500,250)[3000]
\drawline\fermion[\E\REG](19500,300)[3000]
\drawline\fermion[\E\REG](19500,350)[3000]
\drawline\fermion[\E\REG](19500,400)[3000]
\drawline\fermion[\N\REG](3700,1150)[4300]
\drawline\photon[\NW\REG](\particlebackx,\particlebacky)[4]
\drawline\fermion[\E\REG](\particlefrontx,\particlefronty)[3800]
\drawline\fermion[\E\REG](\particlebackx,\particlebacky)[10750]
\drawline\fermion[\E\REG](\particlebackx,\particlebacky)[4100]

\put(4400,1150){\line(2,5){1700}}
\put(4400,1150){\line(1,1){4300}}
\put(4400,1150){\line(5,3){7400}}
\put(4600,1150){\line(5,2){11200}}
\put(16200,5400){\line(2,-5){1740}}
\put(11800,5300){\line(3,-2){6200}}
\put(9000,5300){\line(2,-1){8800}}
\put(6400,5300){\line(5,-2){11200}}
\put(6400,5400){\circle*{600}}
\put(8900,5400){\circle*{600}}
\put(11700,5400){\circle*{600}}
\put(15900,5400){\circle*{600}}

 
\put(6200,6200){${\bf \hat f_1 }$}
\put(7600,6200){${\bf ... }$}
\put(8700,6200){${\bf \hat f_k }$}
\put(11500,6200){${\bf \hat f_{k+1} }$}
\put(14500,6200){${\bf ... }$}
\put(15700,6200){${\bf \hat f_n }$}

\put(500,1000){${\bf {\cal P}_A }$}
\put(2000,8600){${\bf q }$}
\put(2400,3000){${\bf p_1 }$}
\put(4300,4000){${\bf p_2 }$}
\put(3600,6200){${\large\bf _{p_1+q} }$}
\put(9900,2600){${\bf p_{n+1} }$}
\put(8000,1400){${\bf  p_{n+2} }$}
\put(12500,1680){${\bf p'_2 }$}
\put(17500,3200){${\bf p'_{n+1} }$}
\put(12000,-1800){${\bf p_A }$}
\put(20000,6200){${\bf p_f }$}
\put(20000,1000){${\bf {\cal P}_{A-1} }$}
\end{picture}

 

%% file: figure3.tex
\bigphotons
\THICKLINES
{\large 
\begin{picture}(20000,10000)(12000,0)
\drawline\fermion[\E\REG](0,500)[3000]
\drawline\fermion[\E\REG](0,000)[3000]
\drawline\fermion[\E\REG](0,-500)[3000]
\put(3700,00){\circle{1500}}
\drawline\fermion[\NE\REG](4000,700)[3000]
\drawline\photon[\NW\REG](\particlebackx,\particlebacky)[3]
\drawline\fermion[\E\REG](\particlefrontx,\particlefronty)[6000]
\drawline\fermion[\E\REG](4500,000)[7700]
\drawline\fermion[\E\REG](4300,-500)[8000]
\put(10000,-250) {\circle*{700}}

\put(-1800,0)    {${\bf p_{A}}$}
\put(12000,600)  {${\bf p_{p}}$}
\put(7000,600)   {${\bf p_{2}}$}
\put(12000,-1500){${\bf p_{n}}$}
\put(7000,-1500) {${\bf p_{3}}$}
\put(12000,3500) {${\bf p_{f}}$}
\put(5000,4500)  {${\bf q}$}
\put(3750,2000)  {${\bf p_{1}}$}

\put(6500,-5000) {\bf (a)}

\drawline\fermion[\E\REG](16000,500)[3000]
\drawline\fermion[\E\REG](16000,000)[3000]
\drawline\fermion[\E\REG](16000,-500)[3000]
\put(19700,00){\circle{1500}}
\drawline\fermion[\NE\REG](20000,700)[3000]
\drawline\photon[\NW\REG](\particlebackx,\particlebacky)[3]
\drawline\fermion[\E\REG](\particlefrontx,\particlefronty)[6000]
\drawline\fermion[\E\REG](20500,000)[7700]
\drawline\fermion[\E\REG](20300,-500)[8000]
\put(27000,-250){\circle*{700}}
\drawline\fermion[\N\REG](24000,000)[400]
\drawline\fermion[\N\REG](24000,800)[400]
\drawline\fermion[\N\REG](24000,1600)[400]
\drawline\fermion[\N\REG](24000,2400)[400]
\put(28000,600)   {${\bf  p_{p}}$}
\put(22000,600)   {${\bf p_{2}}$}
\put(25000,600)   {${\bf p'_{2}}$}
\put(28000,-1500) {${\bf  p_{n}}$}
\put(23000,-1500) {${\bf p_{3}}$}
\put(28000,3500)  {${\bf p_{f}}$}
\put(22000,3500)  {${\bf  _{p_{1}+q}}$}
\put(21000,4500)  {${\bf q}$}
\put(19750,2000)  {${\bf p_{1}}$}
\put(22500,-5000) {\bf (b)}
\drawline\fermion[\E\REG](32000,500)[3000]
\drawline\fermion[\E\REG](32000,000)[3000]
\drawline\fermion[\E\REG](32000,-500)[3000]
\put(35700,00){\circle{1500}}
\drawline\fermion[\NE\REG](36000,700)[3000]
\drawline\photon[\NW\REG](\particlebackx,\particlebacky)[3]
\drawline\fermion[\E\REG](\particlefrontx,\particlefronty)[6000]
\drawline\fermion[\E\REG](36500,000)[7700]
\drawline\fermion[\E\REG](36300,-500)[8000]
\put(43000,-250){\circle*{700}}
\drawline\fermion[\N\REG](40000,-500)[600]
\drawline\fermion[\N\REG](40000,400)[600]
\drawline\fermion[\N\REG](40000,1400)[600]
\drawline\fermion[\N\REG](40000,2200)[600]
\put(44000,600)     {${\bf  p_{p}  }$}
\put(38200,-1800)   {${\bf p_{3}   }$}
\put(41000,-1800)   {${\bf p'_{3}  }$}
\put(44000,-1800)   {${\bf  p_{n}  }$}
\put(38200,600)     {${\bf p_{2}   }$}
\put(44000,3500)    {${\bf  p_{f}  }$}
\put(38000,3500)    {${\bf  _{p_{1}+q} }$}
\put(37000,4500)    {${\bf q           }$}
\put(35750,2000)    {${\bf p_{1}       }$}
\put(38250,-5000) {\bf (c)}

\drawline\fermion[\E\REG](1000,-14500)[3000]
\drawline\fermion[\E\REG](1000,-15000)[3000]
\drawline\fermion[\E\REG](1000,-15500)[3000]
\put(4700,-15000){\circle{1500}}
\drawline\fermion[\NE\REG](5000,-14300)[3000]
\drawline\photon[\NW\REG](\particlebackx,\particlebacky)[3]
\drawline\fermion[\E\REG](\particlefrontx,\particlefronty)[10000]
\drawline\fermion[\E\REG](5500,-15000)[11700]
\drawline\fermion[\E\REG](5300,-15500)[12000]
\put(15000,-15250){\circle*{700}}
\drawline\fermion[\N\REG](9000,-15000)[400]
\drawline\fermion[\N\REG](9000,-14200)[400]
\drawline\fermion[\N\REG](9000,-13400)[400]
\drawline\fermion[\N\REG](9000,-12600)[400]
\drawline\fermion[\N\REG](12000,-15500)[600]
\drawline\fermion[\N\REG](12000,-14600)[600]
\drawline\fermion[\N\REG](12000,-13600)[600]
\drawline\fermion[\N\REG](12000,-12800)[600]
\put(-800,-15000)    {${\bf p_{A}  }$}
\put(17000,-14400)   {${\bf p_{p}  }$}
\put(13000,-14400)   {${\bf p'_{2} }$}
\put(7000,-14400)    {${\bf p_{2}  }$}
\put(17000,-16800)   {${\bf p_{n}  }$}
\put(13000,-16800)   {${\bf p'_{3} }$}
\put(8000,-16800)    {${\bf p_{3}  }$}
\put(17000,-11500)   {${\bf p_{f}  }$}
\put(6000,-10500)    {${\bf q      }$}
\put(4750,-13000)    {${\bf p_{1}  }$}
\put(10000,-20000) {\bf (d)}
\drawline\fermion[\E\REG](25000,-14500)[3000]
\drawline\fermion[\E\REG](25000,-15000)[3000]
\drawline\fermion[\E\REG](25000,-15500)[3000]
\put(28700,-15000){\circle{1500}}
\drawline\fermion[\NE\REG](29000,-14300)[3000]
\drawline\photon[\NW\REG](\particlebackx,\particlebacky)[3]
\drawline\fermion[\E\REG](\particlefrontx,\particlefronty)[10000]
\drawline\fermion[\E\REG](29500,-15000)[11700]
\drawline\fermion[\E\REG](29300,-15500)[12000]
\put(39000,-15250){\circle*{700}}
\drawline\fermion[\N\REG](36000,-15000)[400]
\drawline\fermion[\N\REG](36000,-14200)[400]
\drawline\fermion[\N\REG](36000,-13400)[400]
\drawline\fermion[\N\REG](36000,-12600)[400]

\drawline\fermion[\N\REG](33000,-15500)[600]
\drawline\fermion[\N\REG](33000,-14600)[600]
\drawline\fermion[\N\REG](33000,-13600)[600]
\drawline\fermion[\N\REG](33000,-12800)[600]
\put(41000,-14400)  {${\bf  p_{p} }$}
\put(37000,-14400)  {${\bf p'_{2} }$}
\put(34000,-14400)  {${\bf p_{2}  }$}
\put(41000,-16800)  {${\bf p_{n}  }$}
\put(37000,-16800)  {${\bf p'_{3} }$}
\put(31000,-16800)  {${\bf p_{3}  }$}
\put(41000,-11500)  {${\bf p_{f}  }$}
\put(30000,-10500)  {${\bf q      }$}
\put(28750,-13000)  {${\bf p_{1}  }$}
\put(34000,-20000)  {\bf (e)}

\end{picture}
}
